\documentclass[twoside]{article}
\usepackage{amsmath}
\usepackage{psfig}

\title{
\Large
{\bf A novel approach to estimate the stability of one-dimensional 
quantum inverse scattering}  
}
\author{ {\bf H.J.S. Dorren} \\
{\small Department of Electrical Engineering, 
Eindhoven University of Technology,
P.O. Box 513,} \\
{\small 5600 MB Eindhoven, The Netherlands} }
\date{}

\begin{document}
\maketitle

\begin{abstract}
We present a novel method to estimate the stability of the Marchenko equation for finite data-sets.
We show that we can derive a recursion relationship for the Fourier 
expansion coefficients 
of the kernel which is solved by the Marchenko equation. The 
method can easily be implemented numerically. Moreover, we discus 
the stability of the one-dimensional inverse scattering problem by using Lyapunov 
exponents. We give conditions on the scattering data to provide stable 
inversion results. A numerical example is given.
\end{abstract}

\vspace{9cm}
\begin{flushleft}
\begin{small}
\hspace{1cm}
E-mail: H.J.S.Dorren@ele.tue.nl 
\end{small}
\end{flushleft}

\newpage

\section{Introduction}

Inverse scattering problems, originally introduced for potential 
scattering in quantum physics (see for instance Ref.\cite{Faddeev} 
and the references therein), have found their 
applications in a wide variety of scientific and technological
fields. An important limitation to the application of inverse
scattering methods to real data is related to the inherently 
unstable behavior of the algorithms involved \cite{Carrion,Koehler}.
We present in this paper a novel method to solve the one-dimensional
inverse scattering problem, and moreover a discussion about the stability 
of the obtained solutions is given.  

The inverse problem for the one-dimensional Schr\"{o}dinger equation 
can be solved by using the Marchenko equation. Burridge showed that the 
inverse problem of the one-dimensional wave equation can be transformed 
into an quantum inverse scattering representation \cite{Burridge}. 
This result made it possible to solve electro-magnetic, optical and acoustic 
inverse problems by using the Marchenko equation. 

The Marchenko equation maps a reflection time series, 
(hereafter to be referred to as the ``data-set''), onto an integral
kernel from which the quantum mechanical potential function can be recovered. 
The data-set is the Fourier transform of the reflection coefficient.
Sabatier has discovered a method to solve
this integral kernel in the Fourier domain for reflection coefficients 
that are a rational function of the wave number \cite{Sabatier}.

In Ref.\cite{Dorren,Dorren1,Dorren2} stability estimates are made for the 
solutions derived by Sabatier.
A disadvantage of the method presented in Ref.\cite{Dorren2} is that 
the method is highly dimensional and therefore not easily implementable 
for realistic data-sets. In this paper we re-open the discussion with 
respect to the solvability and stability of one-dimensional Marchenko-equation 
on the basis of novel results which have recently become available in the field
of the integrability of nonlinear evolution equations \cite{Dorren3,Dorren4}.
It is shown in these publications that large classes of solutions of 
integrable nonlinear evolution equations can be obtained by solving the 
Fourier expansion coefficients of its solution by using a recursion 
relationship. Since it is well known that large classes of nonlinear 
evolution 
equations are integrable by inverse scattering transformations, it is worth 
investigating whether the machinery developed in Ref.\cite{Dorren3,Dorren4} 
can also be applied in solving quantum inverse scattering problems. We 
show in this paper that it is possible to transform the 
Gelfand-Levitan-Marchenko equation into a linear recursion relationship 
which determines the Fourier expansion coefficients of the kernel. 
This result made it also 
possible to investigate the stability of the solutions by using Lyapunov 
exponents which can be computed from the   
data-uncertainties. This implies that we have a simple tool which requires 
only information about the data and data-uncertainties, to investigate 
whether the Marchenko equation is stable.
Secondly, if we might conclude 
that the inverse scattering problem is unstable, we provide machinery to 
extract the largest sub-set from the scattering data which leads to 
a stable solution.    
   
This paper has the following structure: In Section 2 we focus on the 
linearization method which leads to the relationship between
the expansion coefficients of the data and the expansion coefficients of 
the kernel. We firstly investigate the principle for a data-set 
having one Fourier component only, but the method is generalized to data-sets 
having a arbitrary (but finite) number of Fourier components. 
In Section 3, we investigate the stability of the obtained 
solutions by using Lyapunov exponents. A simple 
numerical example is given. The paper is concluded with a discussion. 

\section{The one-dimensional quantum inverse scattering problem}
In this paper, solutions of the quantum inverse scattering problem are discussed. 
The quantum inverse scattering problem is the inverse problem of the
Schr\"{o}dinger equation which is given by: 
\begin{equation}
\psi^{\prime \prime}(k,x) + k^{2} \psi(k,x) = V(x) \psi(k,x)
\label{eq:schrod.2}
\end{equation}
For reasons of simplicity the following restrictions on the scattering
solutions are imposed.
It is assumed that 
$V(x): I \!\! R^{+} \rightarrow I \!\! R $, and
secondly, only incoming waves from the left are taken into account.
The solutions of the Schr\"{o}dinger equation (\ref{eq:schrod.2})
satisfy:
\begin{equation}
\psi(k,x) \sim
\left\{
\begin{array}{ll}
e^{ikx}+R(k)e^{-ikx}     & x \rightarrow - \infty \\
T(k)e^{ikx}              & x \rightarrow + \infty
\end{array}
\right.
\label{eq:psi1.2}
\end{equation}
where $T(k)$ is the transmission coefficient, and $R(k)$ is the
reflection coefficient from the left.
In the following, the data-set is defined by the function $A(t)$.
The relation between the data-set $A(t)$ and the spectral reflection
coefficient $R(k)$ of the Schr\"{o}dinger equation is given by:
\begin{equation}
A(t)= (2 \pi )^{-1}
\int_{- \infty }^{ \infty} \! \!  dk \, R(k)e^{ikt} 
+
\mbox{bound states}
\label{eq:a.2}
\end{equation}
The inverse problem for the Schr\"{o}dinger equation can be solved
using the Marchenko equation:
\begin{equation}
K(x,y)+A(x+y)+ \int_{x}^{\infty} \! \! dz \, K(x,z)A(y+z) = 0
\label{eq:mar.2}
\end{equation}
The potential is recovered from the integral kernel $K(x,y)$ by
\cite{Chadan}:
\begin{equation}
V(x)= - 2 \frac{d}{dx} K(x,x)
\label{eq:v.2}
\end{equation}
In this paper, we firstly discus a method to obtain solutions 
of the Marchenko-equation. The method is based upon results developed
in Ref.\cite{Dorren}.
As a starting point, we assume that the data $A(x)$ can be expanded 
in a Fourier series:
\begin{equation}
A(x) = \sum_{n=1}^{N} A_{n} e^{ik_{n}x}
\label{datfour}
\end{equation}
By considering a data-set which  can be expressed in the form as presented 
in Eq.(\ref{datfour}), we have 
implicitly assumed that the reflection coefficient is a rational 
function of the wave-number. This is not a limitation for the 
application to real data, since 
a measured data-set can be approximated sufficiently by considering a 
series of the form (\ref{datfour}) for which $N \rightarrow \infty$.  
As a first step, we consider the most simple case in which the data-set $A(x)$ 
has only one Fourier component:
\begin{equation}
A(x)= A e^{ikx}
\label{series_dat}
\end{equation}
In this special case, we propose a kernel $K(x,y)$ having the following
structure:
\begin{equation}
K(x,y) =  \sum_{n,m=1}^{\infty} B(n,m) e^{inkx+imky}
\label{series_mod} 
\end{equation}
If the expressions (\ref{series_dat}) and (\ref{series_mod})
are substituted in Eq.(\ref{eq:mar.2}), we obtain the following
result:
\begin{equation}
\sum_{n,m=1}^{\infty} B(n,m) e^{ik(nx+my)} 
+ A e^{ik(x+y)} = -
\sum_{n,m=1}^{\infty} \! \!
A B(n,m) e^{inkx} e^{iky}
\int_{x}^{\infty} \! \! dz \, 
e^{i(m+1)kz}
\label{mar_1}
\end{equation}
If we assume that $k$ is a complex number with positive imaginary part, 
the integral in Eq.(\ref{mar_1}) can be replaced by the primitive of an 
exponential function:
\begin{equation}
\sum_{n,m=1}^{\infty} B(n,m) e^{ik(nx+my)} 
+ A e^{ik(x+y)} = 
\sum_{n,m=1}^{\infty} \! \!
\frac{ A B(n,m) }{i(m+1)k} e^{i(n+m+1)kx} e^{iky}
\label{mar_2}
\end{equation}
We can solve Eq.(\ref{mar_2}) by comparing the exponential functions of equal powers 
on both sides. This is done by firstly defining a 
number $M=n+m$, and hence by comparing all the exponential functions for which 
$M=2,3,4, \cdots$ respectively. If we use the fact that the coefficients 
$A$ and $k$ are determined by the data-spectrum, the kernel $K(x,y)$ can be computed 
by collecting all the coefficients $B(n,m)$. From the $y$-dependence on the 
right-hand side of Eq.(\ref{mar_2}), we can immediately conclude that all
the coefficients $B(n,m)$ vanish for which $m \neq 1$. We consider firstly the 
case that $M=2$:  
\begin{equation}
A e^{ik(x+y)} + B(1,1) e^{ik(x+y)} = 0
\end{equation}
This leads to the result that $B(1,1) = -A$. If we put $M=3$, it can easily be 
verified by substitution that both the coefficients $B(1,2)$ and $B(2,1)$ 
vanish. 
In general, only coefficients $B(n,m)$ for which $m=1$ are 
unequal to zero and follow from the following recursion relationship:
\begin{equation}
B(n+2,1) = A \frac{ B(n,1) }{2ik}
\label{rec_1} 
\end{equation}
By using Eq.(\ref{rec_1}), we can compute the kernel 
$B(n,1)$ as a function of $A$ and $k$:  
\begin{equation}
K(x,y) = - A  e^{ik(x+y)} + \frac{A^{2}}{2ik} e^{3ikx+iky}
- \frac{ A^{3} }{ (2ik)^{2} } e^{5ikx+iky}
+ \cdots
\label{kern_2}
\end{equation}
The potential function $V(x)$ can be recovered from the 
kernel $K(x,y)$ by using the relationship (\ref{eq:v.2}). 
If we substitute $k = i \beta$ and $A=2d$ into 
Eq.(\ref{kern_2}), we obtain the following
expression for $K(x,x)$:
 \begin{equation}
K(x,x) = 2d  e^{-2 \beta x} + \frac{2 d^{2}}{\beta} e^{-4 \beta x}
+ \frac{ 2 d^{3}}{ \beta^{2} } e^{-6 \beta x}
+ \cdots
\label{k_xx2}
\end{equation}
The solution of the kernel (\ref{k_xx2}) is in agreement with solutions
obtained by using Sabatiers approach \cite{Sabatier,Dorren}. 
The advantage of the method presented in this paper is that we 
solve the inverse scattering problem by deriving a linear 
recursion relationship between the Fourier expansion coefficients 
of the data and the expansion coefficients of the kernel.
This implies that we have discovered a method which is easy to implement 
from a computational point of view, and moreover, the stability 
of the obtained solutions can be investigated by using Lyapunov exponents.

In order to generalize this method to the data-set (\ref{series_dat}),
it is also illustrating to discus a data-set consisting of two 
Fourier components:
\begin{equation}
A(x)= A_{1} e^{ik_{1}x} + A_{2} e^{ik_{2}x}
\label{data_3}
\end{equation}
In this special case we propose a kernel $K(x,y)$ having the following 
structure:
\begin{equation}
K(x,y) = \! \! \! \!
\sum_{p,q,r,s =1}^{\infty} 
\! \! \! \!
B(p,q,r,s) e^{ipk_{1}x} e^{iqk_{1} y} 
e^{irk_{2}x} e^{isk_{2} y}
\label{kern_3}
\end{equation}
If we insert Eq.(\ref{data_3}) and Eq.(\ref{kern_3}) in the Marchenko
equation, we obtain the following result:
\begin{equation}
\begin{split}
A_{1} e^{ik_{1}(x+y)} + A_{2} e^{ik_{2}(x+y)} &+
\! \! \! \!
\sum_{p,q,r,s=0}^{\infty}  
\! \! \! \!  
B(p,q,r,s) e^{ipk_{1}x} e^{iqk_{1} y} e^{irk_{2}x} e^{isk_{2} y} \\
&=
\left[
\sum_{p,q,r,s=0}^{\infty} 
\! \!  
\frac{ A_{1} B(p,q,r,s) }{ i ( [q+1] k_{1} + s k_{2} ) }
e^{i(p+q+1)k_{1}x} e^{i(r+s)k_{2}x} e^{i k_{1} y}
\right. \\
&+
\left.
\! \! \! \!
\sum_{p,q,r,s=0}^{\infty} 
\! \! 
\frac{ A_{2} B(p,q,r,s) }{ i ( [s+1] k_{2} + q k_{1} ) }
e^{i(p+q)k_{1}x} e^{i(r+s+1)k_{2}x} e^{i k_{2} y}
\right]
\end{split}
\label{march_3}
\end{equation}
If we use the fact that the coefficients $A_{1}$ and $A_{2}$ are determined 
by the data, we can compute the non-zero coefficients
$B(p,q,r,s)$. This is done, by firstly defining $M=p+q+r+s$ and 
comparing the exponential functions for which $M=2,3, \cdots $ respectively.
If we consider the case that $M=2$, we obtain the following result:
\begin{equation}
A_{1}       e^{ik_{1}(x+y)} + 
A_{2}       e^{ik_{2}(x+y)} +
B(1,1,0,0)  e^{ik_{1}(x+y)} +
B(0,0,1,1)  e^{ik_{2}(x+y)}= 0
\label{step_3}
\end{equation}
From Eq.(\ref{step_3}), we obtain that $B(1,1,0,0) = - A_{1}$ and 
that $B(0,0,1,1) = - A_{2}$. In general, it follows from Eq.(\ref{march_3})
that all the expansion coefficients $B(p,q,r,s)$ can be computed by solving
the following linear iteration relationship:
\begin{equation}
\begin{split} 
B(p+q+1,1,r+s,0) &=  A_{1} \frac{B(p,q,r,s)}{ i([q+1]k_{1} + s k_{2} ) }  \\
B(p+q,0,r+s+1,1) &=  A_{2} \frac{B(p,q,r,s)}{ i([s+1]k_{2} + q k_{1} ) }
\end{split}
\label{it_ser}
\end{equation}
It can be witnessed from Eq.(\ref{it_ser}) that the number of expansion coefficients 
double after each iteration. As an example, we compute the coefficients 
arising from the first three iterations:
\begin{equation}
\begin{array}{c}
B(1,1,0,0) = -A_{1} \\
B(0,0,1,1) = -A_{2}
\end{array}
\hspace{.75cm}
\begin{array}{c}
B(3,1,0,0) = A_{1} \frac{B(1,1,0,0)}{2ik_{1}} \\
B(1,1,2,0) = A_{1} \frac{B(0,0,1,1)}{2ik_{1}} \\
B(2,0,1,1) = A_{2} \frac{B(1,1,0,0)}{2ik_{2}} \\
B(0,0,3,1) = A_{2} \frac{B(0,0,1,1)}{2ik_{2}} \\
\end{array}
\hspace{.75cm}
\begin{array}{c}
B(5,1,0,0) = A_{1} \frac{B(3,1,0,0)}{ 2ik_{1} } \\
B(3,1,2,0) = A_{1} \frac{B(1,1,2,0)}{ 2ik_{1} } \\
B(3,1,2,0) = A_{1} \frac{B(2,0,1,1)}{ i(k_{1}+k_{2}) } \\
B(1,1,4,0) = A_{1} \frac{B(0,0,1,1)}{ i(k_{1}+k_{2}) } \\
B(4,0,1,1) = A_{2} \frac{B(3,1,0,0)}{ i(k_{1}+k_{2}) } \\
B(2,0,3,1) = A_{2} \frac{B(1,1,2,0)}{ i(k_{1}+k_{2}) } \\
B(2,0,3,1) = A_{2} \frac{B(2,0,1,1)}{ 2ik_{2} } \\
B(0,0,5,1) = A_{2} \frac{B(0,0,3,1)}{ 2ik_{2} }
\end{array}
\label{bla}
\end{equation}
It can be witnessed from Eq.(\ref{bla}),
that some of the 
coefficients $B(p,q,r,s)$ occur more than once. Since these coefficients
all belong to the same exponential basis function, we can simply add 
them together. Similarly as for the case that the data-spectrum has only one 
Fourier component, we have a iteration relationship to solve
the coefficients $B(p,q,r,s)$. If we collect all the 
coefficients $B(p,q,r,s)$, we find that the kernel $K(x,y)$ is given 
by the following relationship:
\begin{equation}
K(x,y) = - \sum_{i=1}^{2} 
A_{i}e^{ik_{i}(x+y)} - 
i \sum_{i,j=1}^{2} 
\frac{A_{i}A_{j}}{k_{i}+k_{j}}e^{ik_{j}(x+y)} e^{2ik_{i}x} -
i \sum_{i,j,l=1}^{2} 
\frac{A_{i}A_{j}A_{l}}{(k_{j}+k_{l})(k_{i}+k_{j})}
e^{ik_{l}(x+y)}e^{2i(k_{i}+k_{j})x}
+ \cdots
\label{ker_ser_c}
\end{equation}
This result can be made general by considering the full
data-set $A(x)$ as presented in Eq.(\ref{datfour}).
In this case we propose a kernel $K(x,y)$ having the 
following structure:
\begin{equation}
K(x,y) 
= 
\! \! \! \! \! \! \! \! \! \! \! \!
\sum_{n_{1},\hat{n}_{1} \cdots n_{N},\hat{n}_{N} =1}^{\infty}
\! \! \! \! \! \! \! \! \! \! \! \!
B(n_{1},\hat{n}_{1} \cdots n_{N},\hat{n}_{N})
e^{in_{1}k_{1}x} e^{i\hat{n}_{1}k_{1}z}
\cdots
e^{in_{N}k_{N}x} e^{i\hat{n}_{N}k_{N}z}
\label{kern_gen}
\end{equation}
If we substitute Eq.(\ref{kern_gen}) and Eq.(\ref{datfour}) 
into the Marchenko equation we obtain the following expression:
\begin{equation}
\begin{split}
\sum_{n=1}^{N} A_{n} e^{i k_{n}(x+y)}
&+
\! \! \! \! \! \! \! \! \! \! \! \!
\sum_{n_{1},\hat{n}_{1} \cdots n_{N},\hat{n}_{N} =1}^{\infty}
\! \! \! \! \! \! \! \! \! \! \! \!
B(n_{1},\hat{n}_{1} \cdots n_{N},\hat{n}_{N})
e^{in_{1}k_{1}x} e^{i\hat{n}_{1}k_{1}y}
\cdots
e^{in_{N}k_{N}x} e^{i\hat{n}_{N}k_{N}y} \\
&=
\sum_{n_{1},\hat{n}_{1} \cdots n_{N},\hat{n}_{N} =1}^{\infty}
\left[
\frac{
A_{1} B(n_{1},\hat{n}_{1} \cdots n_{N},\hat{n}_{N})
}{
i([\hat{n}_{1} + 1] k_{1} + \hat{n}_{2} k_{2} 
+ \cdots \hat{n}_{N} k_{N} )
}
e^{ i ( n_{1} + \hat{n}_{1} + 1 ) k_{1} x}
e^{ i ( n_{2} + \hat{n}_{2} ) k_{2} x}
\cdots
e^{ i ( n_{N} + \hat{n}_{N} ) k_{N} x}
e^{ik_{1} y }
\right. \\
&+ 
\cdots
\left.
\frac{
A_{N} B(n_{1},\hat{n}_{1} \cdots n_{N},\hat{n}_{N})
}{
i( \hat{n}_{1} k_{1} + \hat{n}_{2} k_{2} 
+ \cdots [\hat{n}_{N} + 1] k_{N} )
}
e^{ i ( n_{1} + \hat{n}_{1} ) k_{1} x}
e^{ i ( n_{2} + \hat{n}_{2} ) k_{2} x}
\cdots
e^{ i ( n_{N} + \hat{n}_{N} + 1 ) k_{N} x}
e^{ik_{N} y }
\right]
\end{split}
\end{equation}
From this result, we can immediately derive that: 
\begin{equation}
\begin{array}{c}
B(1,1,0,0, \cdots, 0,0) = -A_{1} \\
B(0,0,1,1, \cdots, 0,0) = -A_{2} \\
\vdots \\
B(0,0,0,0, \cdots, 1,1) = -A_{N}
\label{init_fin}
\end{array}
\end{equation}
Moreover, we can derive by using a similar argumentation as used in 
the case that $N=2$ that the recursion relations are given by:
\begin{equation}
\begin{split}
B(n_{1}+\hat{n}_{1}+1,1,n_{2}+\hat{n}_{2},0, \cdots ,n_{N}+\hat{n}_{N},0) 
&=  A_{1}
\frac{
B(n_{1},\hat{n}_{1},n_{2},\hat{n}_{2}, \cdots , n_{N},\hat{n}_{N} ) 
}{ 
i([\hat{n}_{1} + 1] k_{1} + \hat{n}_{2} k_{2} + \cdots \hat{n}_{N} k_{N} ) }    \\
B(n_{1}+\hat{n}_{1},0,n_{2}+\hat{n}_{2}+1,1, \cdots ,n_{N}+\hat{n}_{N},0) &=  A_{2}
\frac{
B(n_{1},\hat{n}_{1},n_{2},\hat{n}_{2}, \cdots , n_{N},\hat{n}_{N} ) 
}{ 
i( \hat{n}_{1} k_{1} + [\hat{n}_{2} +1] k_{2} + \cdots \hat{n}_{N} k_{N} ) }    \\
&\vdots \\
B(n_{1}+\hat{n}_{1},0,n_{2}+\hat{n}_{2},0, \cdots ,n_{N}+\hat{n}_{N}+1,1) &=  A_{N}
\frac{
B(n_{1},\hat{n}_{1},n_{2},\hat{n}_{2}, \cdots , n_{N},\hat{n}_{N} ) 
}{ 
i(\hat{n}_{1}  k_{1} + \hat{n}_{2} k_{2} + \cdots [\hat{n}_{N} +1] k_{N} ) }    \\
\end{split}
\label{it_fin}
\end{equation}
Once the coefficients 
$B(n_{1},\hat{n}_{1},n_{2},\hat{n}_{2}, \cdots , n_{N},\hat{n}_{N} )$ 
are determined, the kernel $K(x,y)$ can be uniquely recovered. Hence the 
kernel $K(x,y)$ can be represented in the following form: 
\begin{equation}
K(x,y) = \sum_{i=1}^{N} 
A_{i}e^{ik_{i}(x+y)} + 
\sum_{i,j=1}^{N} 
\frac{A_{i}A_{j}}{k_{i}+k_{j}}e^{ik_{j}(x+y)} e^{2ik_{i}x} +
\sum_{i,j,l=1}^{N} 
\frac{A_{i}A_{j}A_{l}}{(k_{j}+k_{l})(k_{i}+k_{j})}
e^{ik_{l}(x+y)}e^{2i(k_{i}+k_{j})x}
+ \cdots
\label{ker_fin}
\end{equation} 
The final result which is presented by Eq.(\ref{ker_fin}) is in agreement 
with results obtained in Ref.\cite{Dorren2}. The method 
used for deriving Eq.(\ref{ker_fin}) enables us to investigate
the sensitivity of the method for small data-errors by using Lyapunov exponents. 

Before we proceed, we place a few remarks. 
Firstly, it is interesting that large classes of
solutions of the Marchenko equation can 
be obtained by solving a linear recursion relationship. The instabilities in 
the obtained solutions can be explained by the process of repeated iteration which has 
to be applied to compute the full kernel.
This also implies that the stability of the obtained solutions depends strongly 
on the the numerical value of the parameters  $k_{i}$ and $A_{i}$.
The second remark is that from a conceptual point of view 
there is no difference in solving inverse scattering problems or
solving nonlinear evolution equations \cite{Dorren4}. It is shown this 
publication that both integrable nonlinear evolution equations can be 
linearized using a similar approach as used in this paper. This  
explains why the inverse scattering method works so well in the field of 
integrability.

\section{Stability estimates for the Marchenko equation}
In the previous section solutions of the Marchenko equation 
are discussed. We have shown that we can compute solutions of the 
Marchenko equation by solving the linear recursion relation (\ref{it_fin}). 
Solutions of the Marchenko equation can be unstable due to
process of repeated iteration which is used to compute the expansion coefficients 
of the obtained solutions.
The fact that the Marchenko equation can be solved 
by using a recursion relationship 
enables us to perform the stability analysis of the inverse scattering
problem by using Lyapunov exponents. At a first sight the
bookkeeping for the process of error propagation of the coefficients 
$B(n_{1},\hat{n}_{1},n_{2},\hat{n}_{2}, \cdots , n_{N},\hat{n}_{N} )$ is 
quite tedious, but the analysis can be simplified by considering only
the coefficients which are obtained repeatedly inserting in the same 
recursion relation of Eq.(\ref{it_fin}):
\begin{equation}
\begin{split}
B(n_{1}+2,1,0,0, \cdots ,0,0) &=  A_{1}
\frac{
B(n_{1},1,0,0,\cdots ,0,0) 
}{ i k_{1} [N+1] }    \\
B(0,0,n_{2}+2,1, \cdots ,0,0) &=  A_{2}
\frac{
B(0,0,n_{2},1,\cdots ,0,0) 
}{ i k_{2} [N+1] }       \\
&\vdots \\
B(0,0,0,0, \cdots ,n_{N}+2,1) &=  A_{N}
\frac{
B(0,0,0,0,\cdots, n_{N},1) 
}{ i k_{N} [N+1] }         \\
\end{split}
\label{w_prop}
\end{equation}
By discussing the error-propagation in Eq.(\ref{w_prop}) instead of 
Eq.(\ref{it_fin}) has the advantage that we can investigate the stability
of every pair ($A_{i},k_{i}$) in the data-spectrum.
Suppose that all the
parameters $A_{i}$ and $k_{i}$ in the data lead to coefficients $B$ in Eq.(\ref{w_prop})
which are 
stable for error-propagation. This implies that automatically all the 
other coefficients $B$ (which are computed by solving Eq.(\ref{it_fin}))
are also stable for error-propagation. 
If there are pairs ($A_{i},k_{i}$) leading to instabilities, we can simply  
remove them from the data-set and hence a stable inverse scattering problem 
remains. 
Before computing the Lyapunov exponents, it is convenient to introduce 
a more efficient formulation. We firstly define coefficients 
$B_{i}^{(n)}$ in the following way:
\begin{equation}
B_{i}^{(n)} = B(0,0,0,0,\cdots,n_{i}+2,1,\cdots,0,0)
\end{equation}
Due to imperfections in the data-set, we have to investigate the stability 
of the iteration series (\ref{it_fin}) under the variations 
$A_{i} \rightarrow A_{i}+ \Delta A_{i}$, 
$k_{i} \rightarrow k_{i}+ \Delta k_{i}$ and 
$B_{i}^{(n)} \rightarrow B_{i}^{(n)} + \Delta B_{i}^{(n)}$ respectively. 
If we insert these perturbations in Eq.(\ref{it_fin}), we obtain the 
following result for the perturbation $\Delta B_{i}^{(n)}$:
\begin{equation}
\Delta B_{i}^{(n)} = C_{i} [ \Delta B_{i}^{(n)} ]
\label{stab_1}
\end{equation}
where the operator $C_{i}[ \: \cdot \: ]$ which expresses the error made 
per iteration is given by:  
\begin{equation}
C_{i} [ \: \cdot \: ] =
\frac{ A_{i} }{ i[N+1]k_{i} }  [ \: \cdot \: ] - 
\frac{ 
B_{i}^{(n)} \Delta A_{i}- A_{i} B_{i}^{(n)} [N+1] \Delta k_{i}
}{ 
[N+1]^{2}k_{i}^{2} 
} 
\label{stab_2}
\end{equation}
In principle, after $n$ iterations, we obtain an error which yields:
\begin{equation}
\Delta 
B^{(n+2)}_{i} = C_{i}^{n} [ \Delta A_{i} ]
\label{stab_4}
\end{equation}
If the number of iterations $n$ is large we can re-express 
Eq.(\ref{stab_4}) in the following from:
\begin{equation}
\Delta B^{(n+2)}_{i} = \langle t_{M}^{i} \rangle^{n}_{av} \Delta A_{i} 
\label{stab_5}
\end{equation}    
where the averaged growth of the error per iteration 
$\langle t_{M}^{i} \rangle_{av}$ is given by:
\begin{equation}
\langle t_{M}^{i} \rangle_{av} = 
\frac{
\sum_{j=1}^{M}
|\Delta B^{(j)}_{i} | }{M}
\label{stab_6}
\end{equation}
Eq.(\ref{stab_5}) can be reformulated as:
\begin{equation}
\Delta B^{(n)}_{i} = 
e^{ n \log \langle t_{n}^{i} \rangle_{av} } \Delta A_{i}
= e^{ n \lambda_{n}^{i} } \Delta A_{i}
\label{stab_7}
\end{equation} 
where 
\begin{equation}
\lambda_{n}^{i} = 
\log 
\left\{
\frac{
\sum_{j=1}^{n}
|\Delta B^{(j)}_{i} | }{n}
\right\}
\label{stab_8}
\end{equation}
We have obtained a closed expression for the Lyapunov exponent $\lambda_{n}^{i}$ 
which describes the average growth of errors per iteration. Although the 
the expression for the Lyapunov exponents may look complicated, the method can 
easily be implemented numerically.  
If the Lyapunov exponent is 
positive the errors in the coefficients grow. If the Lyapunov negative
the errors damp out. In the latter case the inverse scattering problem 
is not sensitive for small errors in the data. In principle, we have obtained 
a practical approach to stabilize the inverse scattering
problem since each of the Lyapunov exponents associated with 
Eq.(\ref{stab_8}) corresponds to a single Fourier component of the  
data-spectrum. By removing the Fourier components associated with a positive 
Lyapunov exponent, we can find a stable sub-set of the data-spectrum. 
 
We illustrate this principle numerically for the case that 
that we have only one Fourier component. If we assume that there 
are only uncertainties in $\Delta B$, we obtain that the 
Lyapunov exponent is given by:
\begin{equation}
\lambda^{1} = \lim_{n \rightarrow \infty} \lambda_{n}^{1} =  
\log \left( \frac{d}{2 \beta} \right) 
\label{cond1}
\end{equation}
In Fig.1, the result of the reconstruction is presented for $d=1$ 
and $\beta=10$. It follows from Eq.(\ref{cond1}) that for this choice
of the coefficients $d$ and $\beta$, the Lyapunov exponent is negative 
and the inverse scattering problem is stable.  
If the experiment is repeated with the same value of $\beta$ 
but with $d=0.99$, the curve corresponding to the reconstruction can not be 
distinguished from the unperturbed construction.

This situation changes if we choose $\beta =0.49$ and keep $d$ equal to
unity. The reconstruction is for this case presented by the 
solid line in Fig.2. In this case the reconstruction is according to 
Eq.(\ref{cond1}) sensitive to small changes in $d$.
Indeed, if we solve the inverse scattering problem 
for $d=0.99$, we find that the reconstruction given by the 
broken curve in Fig.2 differs from the unperturbed solution. 

This simple example indicates that the ratio between $d$ and $\beta$ 
determines whether the inverse scattering problem is stable or not. 
The results given in Eq.(\ref{w_prop}) indicate that a similar result
holds in more general cases and forms in principle a tool to investigate which part 
of the data leads to stable inversion results. 
In principle if $\beta \rightarrow 0$, instabilities in the obtained 
solutions can be expected. However in contrast to the results presented in 
Ref.\cite{Dorren}, we show in this paper that there is a delicate balance between 
the amplitude $d$ and frequency $\beta$ of each Fourier component to provide stable 
inversion results. 

\section{Conclusions}
In this paper, we have presented a method to solve the one-dimensional Marchenko
equation for finite data-sets and to investigate the stability of the obtained 
solutions. 
We have established 
a recursion relationship between the Fourier expansion coefficients of the 
data and 
the Fourier expansion coefficients of the kernel. We have also shown that we 
can estimate the stability of the obtained solutions from the
data-uncertainties by using Lyapunov exponents. 

We want to remark that investigating the stability of the Marchenko 
has become an easy task from computational point of view since we have 
presented the Lyapunov exponents in an analytically closed form. 
A practical approach to find a stabilizing transformation for a potentially
unstable data-set consists of simply removing all the unstable pairs 
$(k_{n},A_{n})$ from the data. The low frequency  components in the spectrum
are responsible for the instabilities. Instead of bandpass filtering the 
data-set (to remove the low-frequency components of the data), 
the method presented in this publication enables us to identify the pairs 
$(k_{n},A_{n})$, and their removal leads to  
a more accurate regularization method for the inverse
scattering problem. 

As a concluding remark we have a renewed look on the examples of 
data-contaminations given in Ref.\cite{Dorren}. The first example discussed in
this paper is a small DC-error on the data-set which leads unstable inversion 
results. It is shown  
that a DC-error corresponds with a Fourier component with 
vanishing $k$. As a result of the discussion given in this paper, it can be
understood easily that a vanishing $k$ leads to a positive Lyapunov exponent. 
The other examples discussed in Ref.\cite{Dorren} deal with an amplitude error
and a timing error. It is shown for both examples that if the data-set 
contains Fourier components with small $k$, that inverse scattering problem
is unstable. The results obtained in this paper are in
agreement with Ref.\cite{Dorren}, but additionally quantitative results are given.

%

\newpage
\section*{Captions for Figures}

\noindent
{\bf Figure 1}:
Kernel of $K(x,x)$ is case of one Fourier component as given by 
Eq.(\ref{k_xx2}) for $d=1$ and $\beta=10$ (solid line) and for $d=0.99$ 
and $\beta=10$ (broken curve). In the stable case both curves can
not be distinguished.

\vspace{5mm}
\noindent
{\bf Figure 2}: 
 Kernel of $K(x,x)$ is case of one Fourier component as given by
Eq.(\ref{k_xx2}) for $d=1$ and $\beta=.49$ (solid line) and for $d=0.99$
and $\beta=0.49$ (broken curve). Both curves are scaled with
a factor $10^{-15}$.  

\newpage
\begin{figure}[h]
\centerline{\psfig{figure=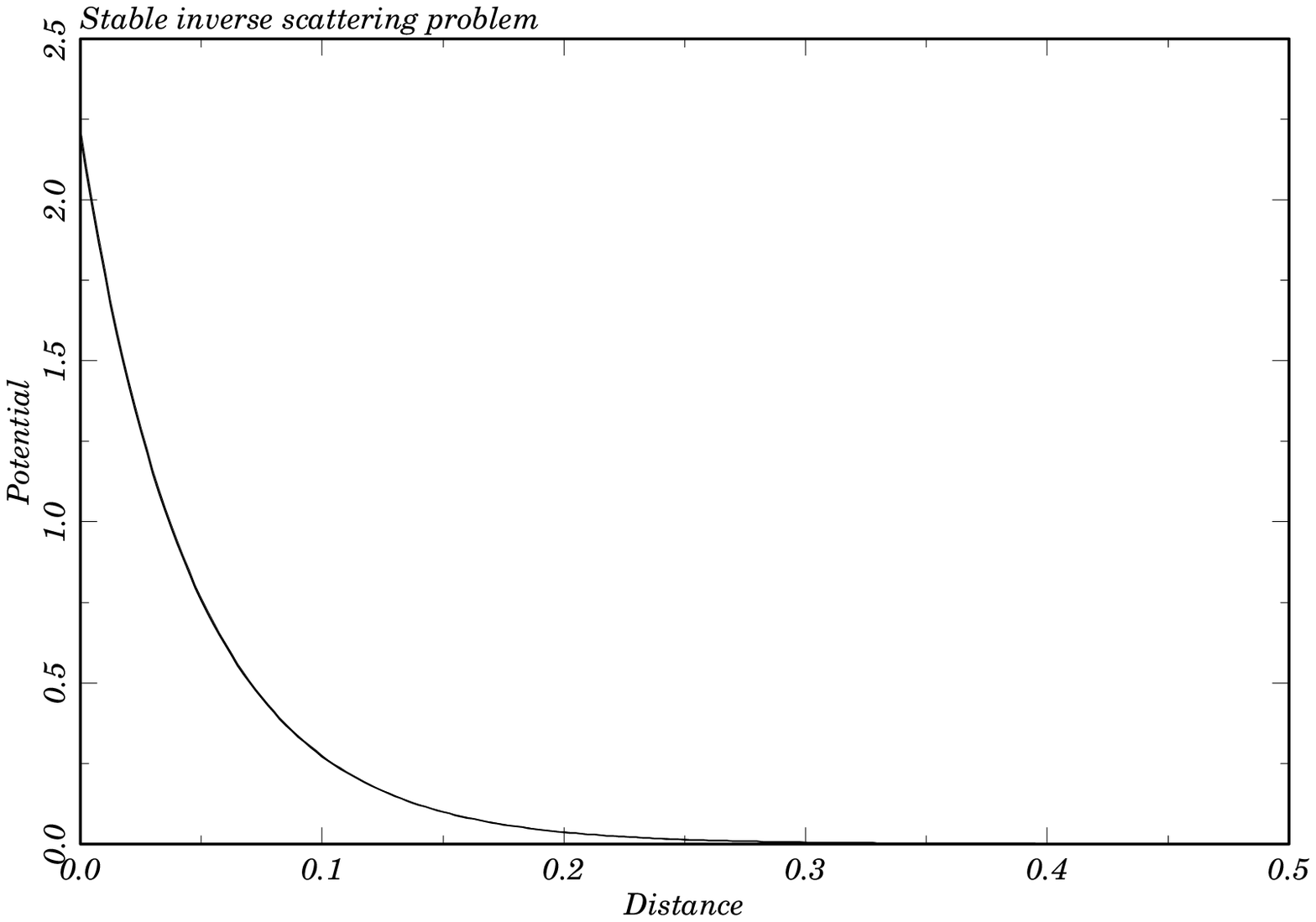,width=15cm,angle=-90}}
\centerline{Figure 1}
\end{figure}

\newpage
\begin{figure}[h]
\centerline{\psfig{figure=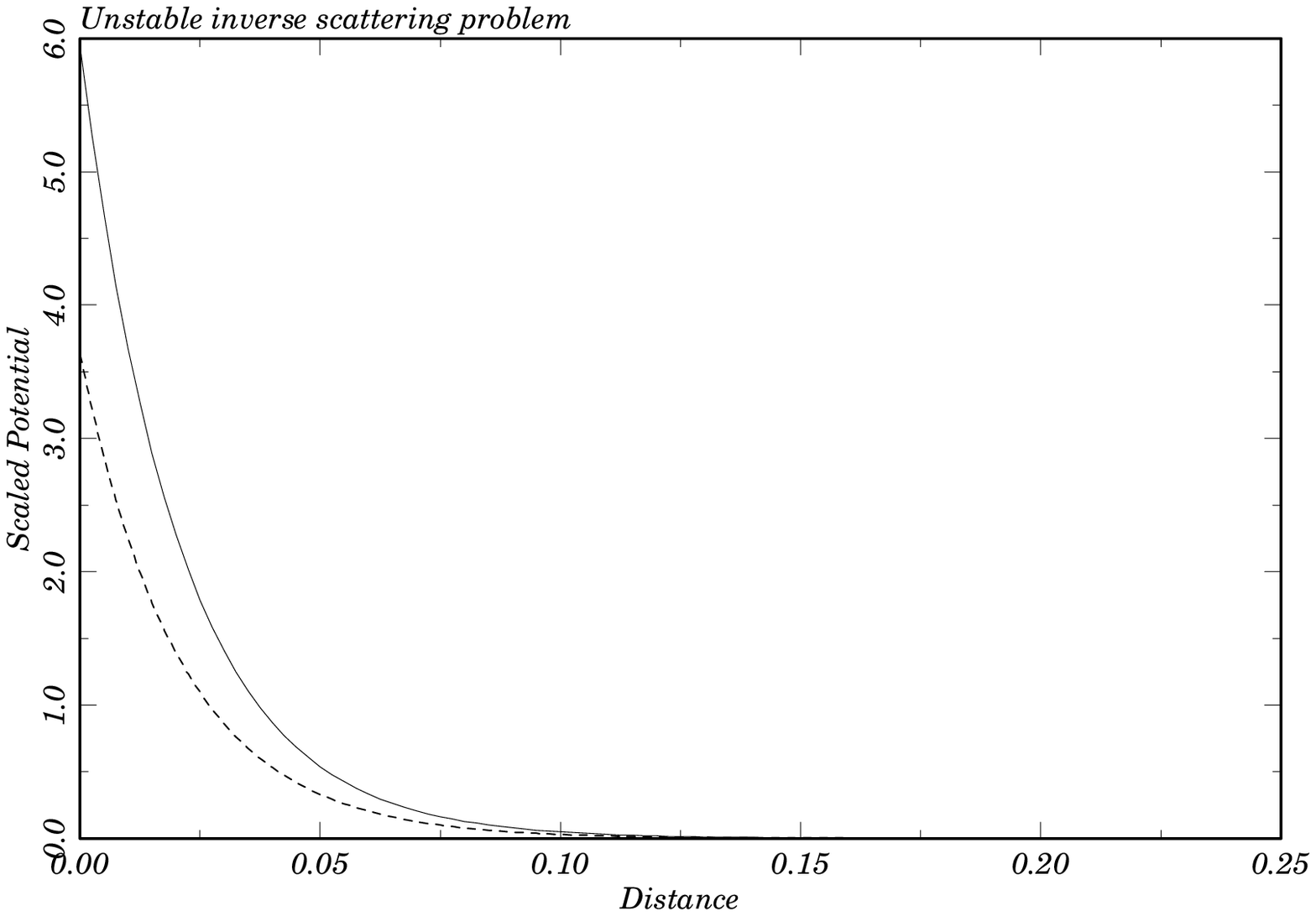,width=15cm,angle=-90}}
\centerline{Figure 2}
\end{figure}


\newpage
\begin{thebibliography}{99}
\bibitem{Faddeev}
L.D. Faddeev, {\em The inverse problem in the quantum theory of
scattering}, J. Math. Phys. {\bf 4}, 1963, 72-104

\bibitem{Carrion}
P.M. Carrion, {\em On the stability of 1D exact inverse methods},
Inverse Problems {\bf 2}, 1986, 1-22.

\bibitem{Koehler}
F. Koehler and M. Taner, {\em Direct and inverse problems relating
reflection coefficients and response for horizontally layered media}, 
Geophysics {\bf 42}, 1977, 1199-1206.

\bibitem{Burridge}
R. Burridge, {\em The Gel'fand-Levitan, The Marchenko, and the 
Gopinath-Sondhi integral equations of inverse scattering theory, 
regarded in the context of inverse impulse-response problems}, Wave
Motion {\bf 2}, 1980, 305-323.

\bibitem{Sabatier}
P.C. Sabatier, {\em Rational reflection coefficients and inverse 
scattering on the line}, Nuovo Cinemento B {\bf 78}, 1983, 235-248.

\bibitem{Dorren}
H.J.S. Dorren, E.J. Muyzert and R.K. Snieder, {\em The stability of one
dimensional inverse scattering}, Inverse Problems {\bf 10}, 865-880, 1994

\bibitem{Dorren1}
H.J.S. Dorren and R.K.snieder, {\em One-dimensional scattering using data
contaminated with errors}, In Lecture Notes on Physics, Quantum Inversion
Theory and Applications, H.V. von Geramb (Ed.), 405-411, Springer-Verlag,
1994. 

\bibitem{Dorren2}
H.J.S. Dorren, and R.K. Snieder, {\em On the stability of inverse problems},
Inverse Problems {\bf 11}, 889-911, 1995.

\bibitem{Dorren3}
H.J.S. Dorren, {\em A linearization method for the Korteweg-de Vries
equation; generalizations to higher dimensional S-integrable
differential equations}, J. Math. Phys. {\bf 39}, 3711-3729, 1998.


\bibitem{Dorren4}
H.J.S. Dorren, {\em On the integrability of nonlinear partial 
differential equations}, preprint (solv-int/9807007).

\bibitem{Chadan}
K.Chadan and P.C. Sabatier, {\em Inverse problems in quantum scattering
theory}, Springer-Verlag, 1989.

\end{thebibliography}
\end{document}